\documentclass[runningheads]{llncs}
\usepackage[T1]{fontenc}
\usepackage{graphicx}
\usepackage{array}

\begin{document}
\title{A Study on Interaction Complexity and Time}
%
%
\author{Leonardo Germán Loza Bonora\inst{1,2}\orcidID{0009-0006-4166-4219} \and
Julián Grigera\inst{1,2,3}\orcidID{0000-0002-7962-4312} \and
Helmut Degen\inst{4}\orcidID{0000-0002-5200-8165}}
\authorrunning{L. Loza Bonora et al.}
%
\institute{LIFIA, Fac. de Informatica, Univ. Nac. de La Plata, La Plata, Argentina \\
\email{\{juliang,lloza\}@lifia.info.unlp.edu.ar} \and
CONICET, Argentina \and
CICPBA, Pcia. de Buenos Aires, Argentina \and
Siemens Corporation, Foundational Technologies, 755 College Road East, Princeton, New Jersey 08540, USA\\
\email{helmut.degen@siemens.com}}
\maketitle              
\begin{abstract}
Testing Web User Interfaces (UIs) requires considerable time and resources, most notably participants for user testing. Additionally, the tests results may demand adjustments on the UI, taking further resources and testing. Early tests can make this process less costly with the help of low fidelity prototypes, but it is difficult to conduct user tests on them, and recruiting participants is still necessary. To tackle this issue, there are tools that can predict UI aspects like interaction time, as the well-known KLM model. Another aspect that can be predicted is complexity, and this was achieved by the Big \textit{I} notation, which can be applied to early UX concepts like lo-fi wireframes. Big \textit{I} assists developers in estimating the interaction complexity, specified as a function of user steps, which are composed of abstracted user actions. Interaction complexity is expressed in mathematical terms, making the comparison of interaction complexities for various UX concepts easy. However, big \textit{I} is not able to predict execution time for user actions, which would be very helpful for early assessment of lo-fi prototypes. To address this shortcoming, in this paper we present a study in which we took measurements from real users (n=100) completing tasks in a fictitious website, in order to derive average times per interaction step. Using these results we were able to study the relationship between interaction complexity and time, and ultimately complement big \textit{I} predictions with time estimates.

\end{abstract}

%
%

\section{Introduction}

In the era of Agile Methods, User Interface (UI) design and development has become hard to integrate. These activities usually require time for evaluating designs, studying users, and testing - which does not always fit into the agile short iterations \cite{Laubheimer2017-AgileUX-misc,Persson2022-UXAgile-InformationSoftwareTechnology-article,Curcio2019-AgileUX-ComputerStandardsInterfaces-article}. In this context, predicting certain qualities of a newly designed UI can be highly beneficial, like efficiency of use. One contributing factor to efficiency is technical response times \cite{Nielsen2017-ResponseTimes-NN-misc}. A 2023 article \cite{Google2023-Performance-misc} reports multiple case studies on how speed matters to users and the business. Another contributing factor is the design of graphical UIs, the focus of this paper. A prominent example of how a design decision increases the efficiency of use is the "Order now" button that Amazon filed as a patent in 1997 \cite{US5960411A}. 
To systematically design for efficiency of use, it would be advantageous to have a technique that can assess this aspect in early UI concepts, such as sketches or low-fidelity wireframes.

Established techniques, such as the Keystroke Level Model (KLM) \cite{Card1980-KLM-Article,CardNewellMoran1983-book}, can predict execution times for experienced users interacting with specific UI elements. However, KLM has known limitations \cite[p. 227]{Olson1990-KLM-HCI-article}, like the need of complete user interface designs, making it challenging to use for early UI concepts. Summative usability tests \cite{ISO25062-2006,RubinChisnellSpool2008-UsabilityTesting-book} measure the effectiveness and efficiency of interactive systems for specified users, goals, and contexts of use, but require an interactive prototype, as well as representatives from the target user groups.

To determine the inherent efficiency of early UI concepts, a new technique is proposed, called the big \textit{I} notation \cite{degen2022big}. It is based on the big \textit{O} notation and allows us to estimate the interaction complexity of earliest UI concepts (e.g., lo-fi wireframes) without user involvement. Interaction complexity is expressed as a mathematical function that helps identifying which parts of a UI concept contribute to high interaction complexity. Big \textit{I}, however, does not determine an execution time which is often desired by stakeholders. 

With this study we aim to determine average interaction time values that can be used to derive an estimated execution time for the interaction complexity, estimated with the big \textit{I} notation. Such times extend the big \textit{I} notation, so interaction complexities and execution times for early UI concepts can be estimated. In this quantitative user study (n=100), we show a validation regarding the relationship between an estimated interaction complexity and measured execution times. With this validation, we intend to show how complexity estimations can also estimate interaction time, which is a widely used measure of efficiency. 

The estimated interaction complexity and an estimated interaction times can be applied to the user interface design of websites and web applications to estimate in an early design phase their efficiency of use. Optimizing the web design for efficiency of use, the success of a website might be improved and can be measured by state-of-the-art website metrics \cite{Freeder2003-webmetrics-PermMeaMetric-article}.

%
%

\section{Background and Related Work}
\label{Sec:Background}
In this section we describe the big \textit{I} notation and its basic components. We also assess similar works in the field that estimate the efficiency of use in any way. 


\subsection{Big \textit{I} notation}
The big \textit{I} notation \cite{degen2022big} ("big \textit{I}") can be used to estimate the interaction complexity of lo-fi and hi-fi UX concepts, production-level user interface design, and user interfaces of launched applications. The letter \textit{I} stands for \textit{interaction}. 

As a motivating example, imagine a building architect designing an airport. While making the first sketches, s/he may want to know the average time it takes for a passenger to walk from a security area to a gate. To do this, the architect measures the distance on the scale drawing with a ruler and transforms it into the real distance based on the scale, then multiply the distance by the average walking speed. The walking time can be determined within minutes, and the architect may decide to modify the design using this information.

Currently, there is no such "UX/HCI ruler", but there are two established techniques that measure time: summative usability tests \cite{RubinChisnellSpool2008-UsabilityTesting-book} and KLM \cite{Card1980-KLM-Article,CardNewellMoran1983-book,Olson1990-KLM-HCI-article}. A usability test requires building a scale model of the airport, recruiting participants, having them walk from a security area to a gate, and measuring the time ('time-on-task'). KLM requires a production-level UI design with all its elements. 
If execution times for certain user actions and UI elements have been measured before they can be reused.
Otherwise, they have to be measured, requiring a similar effort as a summative usability test. Hence, both techniques take significant resources. In real-life projects, there is often no budget and time available to make significant design changes at mature phase in a project.
Big \textit{I} intends to be such a "UX/HCI ruler" by estimating interaction complexity, allowing to make efficiency-related decisions from the earliest stages.
While big \textit{I} is comparable to a ruler, the calculated interaction complexity is comparable to the measured distance. What big \textit{I} currently does not provide is a mapping of the interaction complexity to an average execution time value. This gap should be partially addressed by the study presented in this paper.

The big \textit{I} notation was designed to determine the efficiency of use for UX concepts, spanning from rough sketches or lo-fi wireframes to production-level UI designs or launched applications. Big \textit{I} is particularly useful for estimating efficiency of use for multiple lo-fi wireframes. It expresses the interaction complexity as a mathematical function, making the comparison between various UX concepts easy. Due to the use of mathematical expressions, an estimated interaction complexity has high believability. The big \textit{I} notation is very useful for application domains where the efficiency of use is a critical application property.

To the best of the authors' knowledge, there are multiple properties of big \textit{I} that make it a unique UX/HCI technique. 1) Big \textit{I} does not determine a single value, e.g., a single step or time value, but a complexity function, which allows for easily identifying design decisions that contribute to a relatively high interaction complexity. 2) Big \textit{I} can be applied in the early phases of a design and development process where major design changes can still be made. 3) Its application does not require user involvement. 4) Since big \textit{I} considers abstract user actions, it is modality-agnostic. It can be applied to graphical user interfaces, voice user interfaces, other modalities, and also to multi-modal user interfaces. 5) With some training and practice, the estimated interaction complexity for a given UX concept and a selected user task can be calculated in about an hour, or less. Hence, the application of big \textit{I} is highly cost-effective. The mentioned characteristics of big \textit{I} make it especially compelling for UX practitioners and HCI researchers who have a strong affinity for arithmetic.

Big \textit{I} has some limitations. One of them is that estimated interaction complexity is not precise. Also, it does not determine execution time for user actions. A complexity function is a useful output, but a time value is highly desirable as well. Mapping the estimated interaction complexity to a time value requires determining an average time for interaction steps. The aim of this study is to take time measurements for a selected user task (purchasing a movie ticket) that allows deriving an average execution time per interaction step, so that a time estimate can be determined for an estimated interaction complexity.


\subsection{Related Work}

There are multiple approaches that are similar to the big \textit{I} notation. The keystroke-level model (KLM) predicts execution times for known task scenarios and graphical user interface elements using keyboards, mouse, and monitors as interaction devices \cite{Card1980-KLM-Article,CardNewellMoran1983-book,Olson1990-KLM-HCI-article}. 
There are differences between KLM and big \textit{I}: 1) KLM applies to specific graphical UI elements like text inputs and buttons, requiring a complete user interface design, while big \textit{I} can be applied to both early UI concepts as well as late design artifacts (production-level UIs). 2) KLM determines execution time, while big \textit{I} estimates interaction complexity. 3) KLM is modality specific (graphical UIs), while big \textit{I} is modality agnostic.

The big \textit{I} notation is based on the big \textit{O} notation \cite{Bachmann1894-book,Knuth1976,Landauer1909-book}, but
there are a few key differences between big \textit{I} and big \textit{O}: 1) Big \textit{O} counts the number of instructions executed, treating all types of instructions equally. In contrast, big \textit{I} differentiates between five types of user actions: Think (T), Click a button (C), Scroll a page (S), Enter content (E), and use of an external application (X). This distinction helps to identify design inefficiencies. For instance, if the interaction complexity increases due to the necessary use of an external application (expressed as X), designers can explore alternative UX concepts with the intent to reduce the use of an external application or to completely exclude it. 2) Big \textit{O} simplifies the functional expression of the highest growing function by normalizing its coefficient to 1. In big \textit{I}, the coefficient of the highest growing function is not normalized. Due to the relative slowness of human performance compared to a microprocessor, the coefficient is crucial for determining the execution time of human actions. 3) Big \textit{O} uses a single variable that influences the functional complexity of an algorithm, typically the number of iterations. In big \textit{I}, more than one variable can influence the interaction complexity.

Summative usability test is a technique used to measure effectiveness and efficiency of interactive systems with the involvement of representatives of the target user group. Effectiveness is measured as 'task completion' and efficiency as 'time-on-task' \cite{ISO25062-2006,RubinChisnellSpool2008-UsabilityTesting-book}. It has multiple prerequisites: 1) at least one interactive prototype that behaves similarly to the intended final application for the user tasks under test, 2) representatives of the target user group to perform the test scenarios using the prototype, and 3) time to prepare and execute the usability test, and to analyze and document the results. While summative usability tests have many benefits, they differ from big \textit{I} in several ways: 1) They require an interactive prototype, whereas big \textit{I} can be used on non-interactive UX concepts. 2) Summative usability tests measure time-on-task, which does not reveal directly which parts of a UI cause high complexity and low efficiency. 3) Summative usability tests take significant time, often days or even weeks, while big \textit{I} can be estimated within an hour (with some practice) without user participation.

Overall, the big \textit{I} notation is a novel contribution to the field of human-computer interaction that enables the estimation of the interaction complexity, a unique and useful metric for user interface concepts.   

%
%

\section{Study Design and Preparation}
\label{sec:StudyDesign}

We conducted an online study with participants recruited using the Prolific\footnote{https://www.prolific.com} platform.
The study consisted of a simulated website for online booking of movie tickets\footnote{\url{https://github.com/Keykor/bigi-test-movie-pages}}.
Participants had to complete several tasks, each time booking a ticket under different conditions, using one of 2 different versions (interaction concepts) of the same UI.
On the other hand, we calculated the complexity for each interaction concept, and then compared the completion times.


\subsection{Materials}
\begin{figure}[ht]
\centering
\includegraphics[width=1\textwidth]{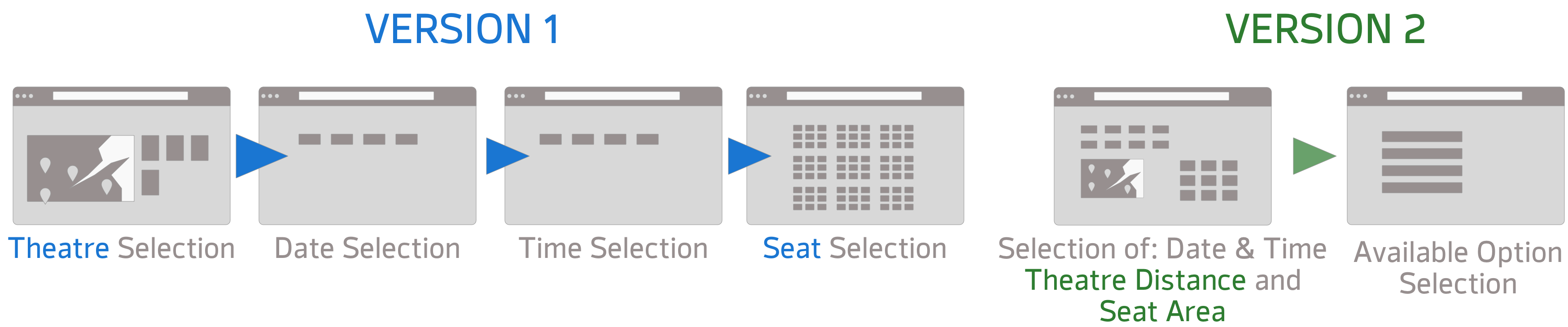}
\caption{Schematics of the 2 versions of the test UI.} \label{fig:versions}
\end{figure}

We selected a movie ticket booking system as our study material. 
The task was to book a ticket for a given movie, date, time, and seat (E12) in two versions of the website (Figure \ref{fig:versions}). 
Version 1 (V1) was a wizard version, where the selection process was step-by-step.
In Version 2 (V2), users first selected a movie, then the criteria (radius of movie theaters, date, time, seat group) and finally the available options on another page.

To collect significant data, we included the necessity to go through multiple task variations for V1, from 1 attempt (seat E12 immediately available) to 5 attempts (trying 5 theatres until E12 was available). 
In the case of V2, selecting the right conditions always guaranteed the preferred seat, so there was no need to try different numbers of attempts, so we created 3 tasks with different conditions (time, theatre distance, etc) for improved validity.
Within each version, tasks were randomized.

We developed a tool that guided the participants through all tasks, providing instructions and conditions for each one that were visible at all times on the right side of the page.
The tasks were provided in random order, in the following way:
\begin{itemize}
    \item the 8 V1 tasks and 3 V2 tasks were provided in batches, but in random order, meaning, a participant could either get all V1 tasks first, or all V2 first.
    \item within each group, the tasks were randomized.
    \item the only fixed variant was a training, single-attempt task for V1.
    This training task was always first within the V1 batch.
\end{itemize}
Additionally, in V1 there were 2 tasks for 1 attempt (considering the training task), and also for 4 and 5 attempts - hence the 8 variants.

Once all 11 tasks were completed, participants were provided with a link to go back to the Prolific platform and get their tasks approved.
The tool captured their selections and low-level events such as mouse moves and clicks, which also allowed us to calculate durations at the interaction step level.


\subsection{Estimating the interaction complexity}
\label{sec:Material:bigI}
We calculated interaction complexity for both versions of the UI using big \textit{I}.
This is estimated by following these steps: 1) Determine variables, 2) Determine and calculate interaction steps, 3) Sum up interaction steps, 4) Normalize interaction steps, 5) Simplify interaction steps, and 6) Instantiate interaction steps. 

The big \textit{I} notation uses the following abstract user actions for the estimate: 

\begin{itemize}
    \item T: Think
    \item E: Enter content
    \item C: Click buttons
    \item S: Scroll
    \item X: Use of external application or tool
\end{itemize}


\paragraph{Step 1: Determine variables}

We identify the variables that influence interaction complexity. These variables apply to both versions, but not all variables might be relevant for both versions. For each abstract user action, we determine how many user actions are performed per user step. If the number of user actions depend on a variable, the variable will be defined first. In some cases, the number of user actions depends on a constant. 
The variables are:
\begin{itemize}
    \item m: number of displayed movies
    \item r: number of radius for theater selection
    \item t: number of displayed movie theaters (for version 1 only)
    \item d: number of dates
    \item s: number of time slots
    \item g: number of seat groups (for version 2 only)
    \item a: number of attempts to find the show with the right seat (for version 1 only)
    \item o: number of options (for version 2 only)
\end{itemize}

\subsubsection{Version 1}

\paragraph{Step 2: Determine and calculate interaction steps}
We identify the different user steps and express the complexity per user step by identifying the user actions (i.e., T,C,E,S,X) and how often they are performed. Introduced variables are used for some of the user steps. 

\begin{itemize}
    \item User step 1: Review $m$ movies, select one, and go to next page: 
    
    $f_1(m,r,t,d,s,g,a,o) = m \cdot T + E + C$
    \item User step 2: Review $r$ theater radius, select one, and go to next page (repeated $a$ times): $f_2(m,r,t,d,s,g,a,o) = a \cdot (r \cdot T + E + C)$
    \item User step 3: Review $t$ theaters, select one, and go to next page (repeated $a$ times): $f_3(m,r,t,d,s,g,a,o) = a \cdot (t \cdot T + E + C)$
    \item User step 4: Review $d$ dates, select one, and go to next page (repeated $a$ times): $f_4(m,r,t,d,s,g,a,o) = a \cdot (d \cdot T + E + C)$
    \item User step 5: Review $s$ time slots, select one, and go to next page (repeated $a$ times):
    $f_5(m,r,t,d,s,g,a,o) = a \cdot (s \cdot T + E + C)$
    \item User step 6: Review E12 seat (repeated $a$ times): $f_6(m,r,t,d,s,g,a,o) = a \cdot T$
    \item User step 7: Go back to theater selection (repeated $a - 1$ times): 
    
    $f_7(m,r,t,d,s,g,a,o) = 3 \cdot (a - 1) \cdot C$ (if seat E12 is not available)
    \item User step 8: Select seat E12 and go to next page: $f_8(m,r,t,d,s,g,a,o) = E + C$ (if seat E12 is available)
    \item User step 9: Review and confirm selection: $f_9(m,r,t,d,s,g,a,o) = T + C$
\end{itemize}

Due to the unavailability of the seat E12, steps 2 through 6 are performed $a$ times. Step 7 is performed if seat E12 is unavailable and performed $a-1$ times. Step 8 is performed if seat E12 is available. The user action of selecting an item is expressed as $E$. The user action of going to the next or previous page is expressed as $C$. The user action of reviewing the items and making a decision is expressed as $T$. 

\paragraph{Step 3: Sum up interaction steps}
In this step, the sum of all introduced user steps is calculated. The sum is organized by user actions. In our case, the user actions T and C are used.  

$f_s(m,r,t,d,s,g,a,o) = \sum_{i=1}^{9} f_i(m,r,t,d,s,g,a,o)$ = $(m \cdot T + E + C) + (a \cdot (r \cdot T + E + C)) + (a \cdot (t \cdot T + E + C)) + (a \cdot (d \cdot T + E + C)) + (a  \cdot (s \cdot T + E + C)) + T + (3 \cdot (a - 1) \cdot C) + (E + C) + (T + C)$ =
$(m + 2 + a \cdot (r + t + d + s)) \cdot T + (4 \cdot a + 2) \cdot E + (7 \cdot a + 1) \cdot C$ 

\paragraph{Step 4: Normalize interaction steps}
In this step, the user actions (T and C) will be replaced by abstract \textit{Interaction Steps (IS)}.

$f_{IS}(m,r,t,d,s,g,a,o) = (m + 2 + a \cdot (r + t + d + s)) \cdot IS + (4 \cdot a + 2) \cdot IS + (7 \cdot a + 1) \cdot IS  = m + 5 + (a \cdot (r + t + d + s + 11)) \cdot IS$

\paragraph{Step 5: Simplify interaction steps}
In this step, the highest growing function is kept, including its coefficient. The lower growing functions are removed. The result is the "estimated interaction complexity." Its size will be classified.  

$I(g_{IS}(m,r,t,d,s,g,a,o) = I(m + 5 + (a \cdot (r + t + d + s + 11)) = I(a \cdot (r + t + d + s + 11))$ ("quadratic interaction complexity")

\paragraph{Step 6: Instantiate interaction steps}
Building on the outcome of step 4 "Normalize..." the actual number of interaction steps is calculated. A number is assigned to each variable, which allows calculating the total number of interaction steps.

$f_{IS}(m=6,r=4,t=7,d=4,s=6,g=0,a=5,o=0) = (6 + 5 + 5 \cdot (4 + 7 + 4 + 6 + 11)) \cdot$ IS $= 11 + 5 \cdot 32$ IS = 171 IS

\subsubsection{Version 2}

\paragraph{Step 2: Determine and calculate interaction steps}

\begin{itemize}
    \item User step 1: Review $m$ movies, select one and go to next page: 
    
    $f_1(m,r,t,d,s,g,a,o) = m \cdot T + E + C$
    \item User step 2: Review $r$ theater radius and select one, review $d$ dates and select one, reviw $s$ time slots and select one, review $g$ seat groups and select one, and go to next page: $f_2(m,r,t,d,s,g,a,o) = (r \cdot T + E) + (d \cdot T + E) + (s \cdot T + E) + (g \cdot T + E) + C = (r + d + s + g) \cdot T + 4 \cdot E$ + C
    \item User step 3: Review $o$ options, select one and go to next page: 
    
    $f_3(m,r,t,d,s,g,a,o) = o \cdot T + E + C$
    \item User step 4: Review and confirm selection: $f_4(m,r,t,d,s,g,a,o) = T + C$
\end{itemize}

\paragraph{Step 3: Sum up interaction steps}

$f_s(m,r,t,d,s,g,a,o) = \sum_{i=1}^{4} f_i(m,t,r,d,s,a,o)$ = $(m \cdot T + E + C) + (r \cdot T + E + d \cdot T + E + s \cdot T + E + g \cdot T + E + C) + (o \cdot T + E + C) + (T + C)$ = $(m + r + d + s + g + o + 1) \cdot T + 7 \cdot E + 4 \cdot C$   

\paragraph{Step 4: Normalize interaction steps}

$f_{IS}(m,r,t,d,s,g,a,o) = (m + r + d + s + g + o + 1) \cdot IS + 7 \cdot IS + 4 \cdot IS = (m + r + d + s + g + o + 12) \cdot IS$

\paragraph{Step 5: Simplify interaction steps}

$I(g_{IS}(m,r,t,d,s,g,a,o) = I(m + r + d + s + g + o + 12) = I(m + r + d + s + g + o)$ ("linear interaction complexity")

\paragraph{Step 6: Instantiate interaction steps}

$f_{IS}(m=6,r=4,t=0,d=4,s=4,g=9,a=0,o=7) = (6 + 4 + 4 + 4 + 9 + 7 + 12) \cdot IS = 46 \cdot IS$

The question is, what are average times for users to perform the estimated user steps, so an execution time for an estimated interaction complexity can be estimated.


\subsection{Calculating execution time with KLM}
\label{sec:Material:KLM}

In order to relate the measured execution times in this research with state-of-the-art, we calculate the KLM execution times for the same user steps. 
KLM uses operators with average execution time values \cite[Figure 7 and p. 241]{Olson1990-KLM-HCI-article}. 

\begin{itemize}
    \item K: Enter single keystroke: 0.23 sec (not applicable)
    \item M: Point with a mouse to a target: 1.5 sec
    \item C: Click and release the mouse: 0.23 sec
    \item \textbf{T: M + C = 1.5 sec + 0.23 sec = 1.73 sec}
    \item S: Make a saccade: 0.23 sec
    \item P: Perceive: 0.1 sec
    \item R: Retrieve from long-term memory to working memory: 1.2 sec (not applicable)
    \item E: Execute mental step: 0.07 sec
    \item \textbf{Q: S + P + E = 0.23 sec + 0.1 sec + 0.07 sec = 0.4 sec}
\end{itemize}

Since there is no text input, $K$ is not used for this study. There is also no access to the long-term memory because information is new for participants, excluding R from the calculation for this study. For our calculation, only $T$ and $Q$ are used.


\subsubsection{Version 1}

\paragraph{Step 1: Calculate the execution time per user step}
\begin{itemize}
    \item User step 1: Review $m$ movies, select one, and go to next page: 
    
    $KLM_1(m,t,r,d,s,a) = m \cdot Q + T + T$
    \item User step 2: Review $r$ theater radius, select one, and go to next page (repeated $a$ times): $KLM_2(m,t,r,d,s,a) = a \cdot (r \cdot Q + T + T)$
    \item User step 3: Review $t$ theaters, select one, and go to next page (repeated $a$ times): $KLM_3(m,t,r,d,s,a) = a \cdot (t \cdot Q + T + T)$
    \item User step 4: Review $d$ dates, select one, and go to next page (repeated $a$ times): $KLM_4(m,t,r,d,s,a) = a \cdot (d \cdot Q + T + T)$
    \item User step 5: Review $s$ time slots, select one, and go to next page (repeated $a$ times): $KLM_5(m,t,r,d,s,a) = a \cdot (s \cdot Q + T + T)$
    \item User step 6: Review E12 seat (repeated $a$ times): $KLM_6(m,t,r,d,s,a) = a \cdot Q$
    \item user step 7: Go back $a - 1$ times to theater selection (repeated $a - 1$ times): $KLM_7(m,t,r,d,s,a) = (a - 1) \cdot (Q + T)$ (if seat E12 is not available)
    \item User step 8: Select seat E12 and go to next page: $KLM_8(m,t,r,d,s,a) = T + T$ (if seat E12 is available)
    \item User step 9: Review and confirm selected ticket: $KLM_9(m,t,r,d,s,a) = Q + T$
\end{itemize}

\paragraph{Step 2: Sum up the KLM steps}
$KLM_S(m,r,t,d,s,a) = \sum_{i=1}^{9} KLM_i(m,t,r,d,s,a)$ = $(m \cdot Q + T + T) + (a \cdot (r \cdot Q + T + T)) + (a \cdot (t \cdot Q + T + T)) + (a \cdot (d \cdot Q + T + T)) + (a \cdot (s \cdot Q + T + T)) + a \cdot Q + ((a - 1) \cdot (Q + T)) + (T + T) + (Q + T)$ = $(m + a \cdot (r + t + d + s + 2)) \cdot Q + (4 + 8\cdot a) \cdot T$

\paragraph{Step 3: Instantiate the KLM steps with execution time}
$KLM_S(m,r,t,d,s,a)$ = (m + a $\cdot$ (r + t + d + s + 2)) $\cdot$ 0.4 sec + (4 + 8 $\cdot$ a) $\cdot$ 1.73 sec

\paragraph{Step 4: Instantiate the variables}
$KLM_S(m=6,r=4,t=7,d=6,s=5,a=5)$ = (6 + 5 $\cdot$ (4 + 7 + 6 + 5 + 2)) $\cdot$ 0.4 sec + (4 + 8 $\cdot$ 5) $\cdot$ 1.73 = 126 $\cdot$ 0.4 sec + 44 $\cdot$ 1.73 sec = 50.4 sec + 76.12 sec = 126.52 sec.


\subsubsection{Version 2}

\paragraph{Step 1: Calculate the execution time per user step}
\begin{itemize}
    \item User step 1: Review $m$ movies, select one and go to next page:\\
    $KLM_1(m,r,t,d,s,o) = m \cdot Q + T + T$
    \item User step 2: Review $r$ theater radius and select one, review $d$ dates and select one, review $s$ time slots and select one, review $g$ seat groups and select one, and go to next page: $KLM_2(m,r,t,d,s,o) = (r \cdot Q + T) + (t \cdot Q + T) + (d \cdot Q + T) + (s \cdot Q + T) + Q + T$
    \item User step 3: Review $o$ options, select one and go to next page:\\
    $KLM_3(m,r,t,d,s,o) = o \cdot Q + T + T$
    \item User step 4: Review and confirm selection: $KLM_4(m,r,t,d,s,o) = Q + T$
\end{itemize}

\paragraph{Step 2: Sum up the KLM steps}
$KLM_S(m,r,t,d,s,o)$ = $\sum_{i=1}^{4} KLM_i(m,r,t,d,s,o)$ = (m $\cdot$ Q + T + T) + s(r $\cdot$ Q + T) + (t $\cdot$ Q + T) + (d $\cdot$ Q + T) + (s $\cdot$ Q + T) + Q + T) + (o $\cdot$ Q + T + T) + (Q + T) = (m + r + t + d + s + o + 2) $\cdot$ Q + 9 $\cdot$ T

\paragraph{Step 3: Instantiate the KLM steps}
$KLM_S(m,r,t,d,s,o)$ = (m + r + t + d + s + o + 2) $\cdot$ 0.4 sec + 9 $\cdot$ 1.73 sec

\paragraph{Step 4: Instantiate the variables}

$KLM_S(m=6,r=4,t=7,d=6,s=5,o=5)$ = (6 + 4 + 7 + 6 + 5 + 5 + 2) $\cdot$ 0.4 sec + 9 $\cdot$ 1.73 sec = 35 $\cdot$ 0.4 sec + 9 $\cdot$ 1.73 sec = 14.00 sec + 15.57 sec = 29.57 sec


\subsection{Participants and Execution}
We conducted our study using the Prolific platform, a marketplace for recruiting users to run online tasks and surveys.
We required fluent English speakers, with 98/100 approved tasks.
We also required the tasks to be run on a desktop device, but this was not technically possible to enforce.
We published an estimated time of 30' that we established by conducting prior in-person test runs. Each participant received 3.00 GBP.
Note that participants were allowed take more time than this estimation.

We added 2 questions at the start of the tasks: how frequently they bought online movie tickets (never/about once a year/many times a year), and how frequently they made online purchases in general (never/about once a month/many times a month).
The different time spans (year / month) is intentional, as we expect movie ticket purchases to be less frequent than general purchases.

The execution took place in 3 sessions between February 3rd and February 7th, 2025.
We recruited a total of 100 participants in two stages: a first dry-run of 10 participants to adjust potential capture issues, and the remaining 90 at a later stage.
An extra 12 participants were not accounted for: 10 abandoned the task and 2 timed out.
5 participants did not complete all the tasks, but their results were still useful since our analysis was not session-wide.

%
%

\section{Results}
\label{Sec:Results}

The captured data can be organized in 4 levels: session, task, page visit, and interaction step. This is important to understand the data structure but only 2 levels, task and interaction step, are of interest for the study.

Table \ref{tab:times_for_iterations} shows the completion times (min / max / mean) for each task.
In V1 tasks, it should be noted that there were 2 versions for both 4-attempt and 5-attempt tasks.
In this case they were reported both separately and combined.
The same happens for V2, since there are 3 tasks but there is no specified number of attempts - so all 3 tasks are equivalent in this sense.
The training task is reported separately, in the first line of the table. The rightmost 3 columns of the table show interaction speed in Interaction Steps per second as a unit.

We removed outliers using the interquartile range method, which consists of defining an interquartile range (IQR) between Q1 and Q3.
The lower limit was defined as (Q1 - 1.5 * IQR) and the upper as (Q3 + 1.5 * IQR). 
Notice that even if we obtained a total of 100 sessions (i.e. participants) the reported \textbf{n} per task varies, since the removal of outliers was done per-task.

\begin{table}
    \caption{Interaction speeds for version 1 and version 2}
        \begin{tabular}{
        >{\raggedright\arraybackslash}m{3cm}
        >{\raggedright\arraybackslash}m{1cm} |
        >{\centering\arraybackslash}m{1.3cm}
        >{\centering\arraybackslash}m{1.3cm}
        >{\centering\arraybackslash}m{1.3cm} |
        >{\centering\arraybackslash}m{1.3cm} 
        >{\centering\arraybackslash}m{1.3cm} 
        >{\centering\arraybackslash}m{1.3cm}} 
        Version (Number of attempts)    & IS              & Min (sec) &  Max (sec)  & Mean (sec)     & Max (IS/sec) & Min (IS/sec) & Mean (IS/sec) \\
        \hline
        V1 (1)* (n=74)	&	43	&	16.54	&	197.73	&	80.50	& 2.60    & 0.22  &	0.53 \\
        \hline
        V1 (1) (n=85)	& 43  &	13.88	& 238.76	& 42.20	& 3.10 & 0.18 &	1.02 \\
        V1 (2) (n=86)	& 75  &	19.54	& 226.99	& 65.45	& 3.84 & 0.33 &	1.15 \\
        V1 (3) (n=84)	& 107 &	23.57	& 217.44	& 90.22	& 4.54 & 0.49 &	1.19 \\
        V1 (4) (n=165) &	139	& 17.12	&	237.31	&	98.64	& 8.12    & 0.59  &	1.41 \\
        V1 (5) (n=158) &	171	& 20.98	&	238.26	&	118.92	& 8.15    & 0.72  &	1.44 \\
        \hline
        V2 (1) (n=260)&	46	&	13.06	&	232.65	&	70.15	& 3.52  & 0.20  &	0.66 \\
        \hline
        Mean (n = 912)  & -    & -         & -  &   -       & 8.15    & 0.18  & 1.05\\ 
        \end{tabular}
        *Training session
    \label{tab:times_for_iterations}
\end{table}

To express interaction steps per time, we introduce the term \textit{interaction speed}. The interaction speed is expressed as interaction steps per time (IS/sec).

\begin{figure}
    \centering
    \includegraphics[width=0.8\linewidth]{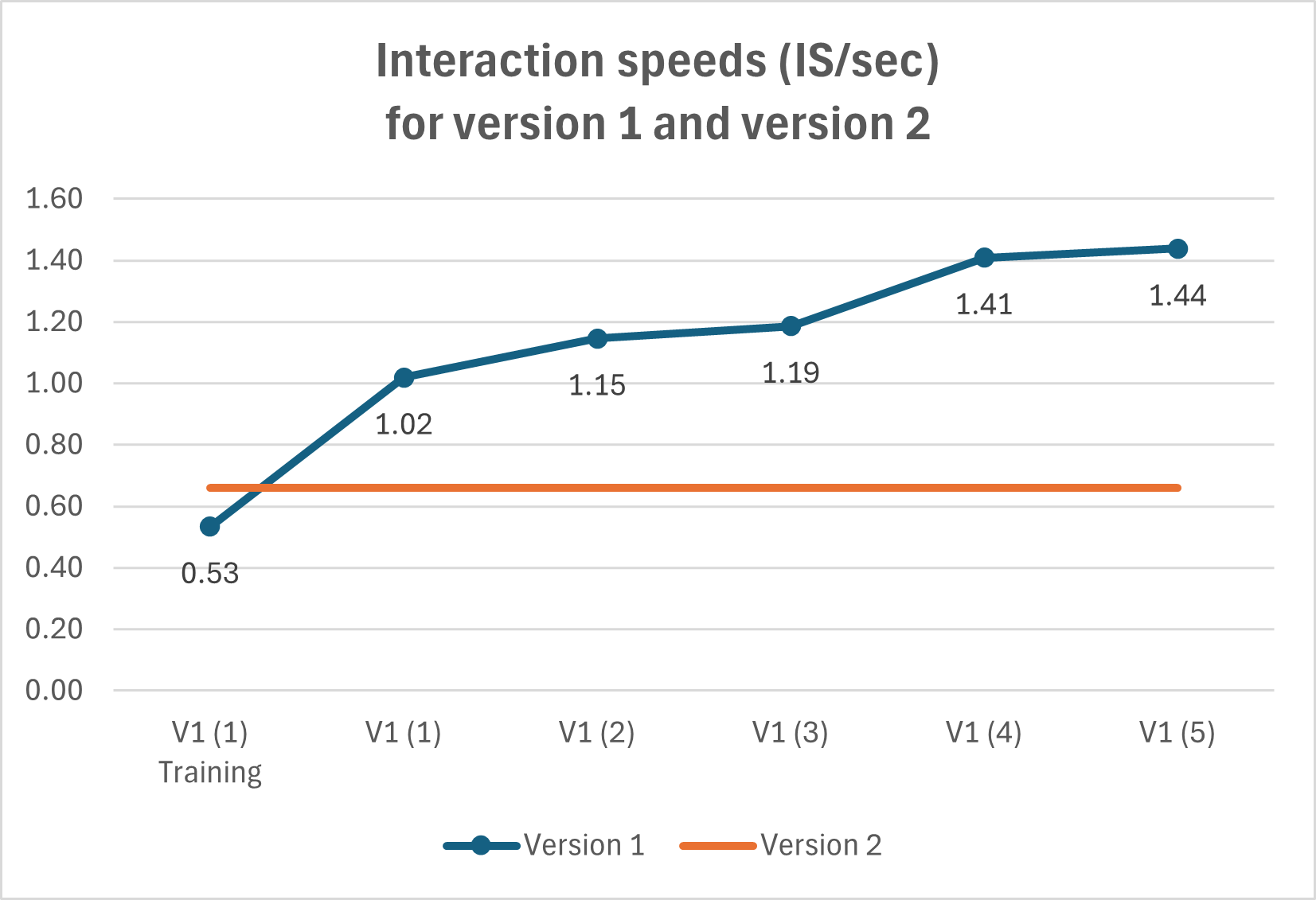}
    \caption{Interaction speeds for version 1 and version 2}
    \label{fig:InteractionPerformance}
\end{figure}

The results of the measured interaction speeds for version 1 and version 2 are depicted in Table~\ref{tab:times_for_iterations}. The resulting mean interaction speed for version 1 is 1.20 IS/sec (not shown in Table~\ref{tab:times_for_iterations}) and the mean interaction speed for version 2 is 0.66 IS/sec. The mean interaction speed for version 1 and version 2 is 1.05 IS/sec, ranging from 0.18 IS/sec at the lower end up to 8.15 IS/sec at the upper end (see also Figure~\ref{fig:InteractionPerformance}).

The interaction speeds for individual user steps for version 1 is depicted in Table~\ref{tab:TimesperAction1} and for version 2 in Table~\ref{tab:TimesperAction.2}.

\begin{table}
    \centering
    \caption{Interaction speeds for version 1}
    \label{tab:TimesperAction1}
    \begin{tabular}{
        >{\raggedright\arraybackslash}m{5.8cm}   
        >{\centering\arraybackslash}m{0.4cm}     
        |>{\centering\arraybackslash}m{0.8cm}  
        >{\centering\arraybackslash}m{0.8cm}  
        >{\centering\arraybackslash}m{0.8cm}  
        |>{\centering\arraybackslash}m{1.1cm}  
        >{\centering\arraybackslash}m{1.1cm}  
        >{\centering\arraybackslash}m{1.1cm}  
    }
         User steps & IS & Min (sec) & Max (sec) & Mean (sec) & Max (IS/sec) & Min (IS/sec) & Mean (IS/sec) \\
        \hline
        1a: View six movies and select one (n=676) & 7 & 1.17 & 12.70 & 4.75 & 0.55 & 5.98 & 1.47 \\
        1b: Click next to radius theater map (n=701) & 1 & 0.42 & 3.70 & 1.59 & 0.27 & 2.38 & 0.63 \\
        2: View four distance filters and select one (n=432) & 5 & 1.19 & 10.16 & 3.97 & 0.49 & 4.20 & 1.26 \\
        3a: View six movie theaters and select one (n=2078) & 7 & 0.50 & 10.35 & 3.27 & 0.68 & 14.00 & 2.14 \\
        3b: Click next to date (n=2161) & 1 & 0.29 & 5.51 & 2.03 & 0.18 & 3.45 & 0.49 \\
        4a: View six dates and select one (n=2255) & 7 & 0.51 & 4.46 & 1.92 & 1.57 & 13.73 & 3.65 \\
        4b: Click next to show (n=2226) & 1 & 0.21 & 3.77 & 1.57 & 0.27 & 4.76 & 0.64 \\
        5a: View four times and select one (n=2439) & 5 & 0.17 & 4.94 & 1.94 & 1.01 & 29.41 & 2.58 \\
        5b: Click next to seat selection (n=2460) & 1 & 0.23 & 3.23 & 1.40 & 0.31 & 4.35 & 0.71 \\
        6: View and decide to select a seat (n=648) & 2 & 0.39 & 4.38 & 1.72 & 0.46 & 5.13 & 1.16 \\
        7a: Click back to show (n=1763) & 1 & 0.71 & 9.90 & 3.46 & 0.10 & 1.41 & 0.29 \\
        7b: Click back to date (n=1542) & 1 & 0.06 & 3.17 & 1.22 & 0.32 & 16.67 & 0.82 \\
        7c: Click back to theater (n=1525) & 1 & 0.01 & 2.95 & 1.08 & 0.34 & 100.00 & 0.93 \\
        8: Click next to summary (n=651) & 1 & 0.57 & 8.00 & 2.80 & 0.13 & 1.75 & 0.36 \\
        9a: View summary and decide to buy (n=652) & 2 & 1.32 & 14.29 & 5.29 & 0.14 & 1.52 & 0.38 \\
        9b: Click back to movies (n=69) & 1 & 0.98 & 25.76 & 7.04 & 0.04 & 1.02 & 0.14 \\
        \hline
    \end{tabular}
\end{table}

\begin{table}
    \centering
    \caption{Interaction speeds for version 2}
    \label{tab:TimesperAction.2}
    \begin{tabular}{>{\raggedright\arraybackslash}m{5.4cm} 
    >{\centering\arraybackslash}m{0.3cm} |
    >{\centering\arraybackslash}m{0.8cm} 
    >{\centering\arraybackslash}m{0.8cm} 
    >{\centering\arraybackslash}m{0.8cm} | 
    >{\centering\arraybackslash}m{1.1cm} 
    >{\centering\arraybackslash}m{1.1cm} 
    >{\centering\arraybackslash}m{1.1cm}} 
    User steps                                  & IS             & Min (sec) & Max (sec) & Mean (sec) & Max (IS/sec) & Min (IS/sec) & Mean (IS/sec)\\
     \hline
        1a: View six (m) movies and select one (n=256)	&   7  &  1.50  &  17.41  &  5.70  &  4.66  &  0.40  &  1.23 \\ 
        1b: Click next to date page (n=261)	&   1  &  0.29  &  4.68  &  1.67  &  3.42  &  0.21  &  0.60 \\ 
        2a: View six (d) dates and select one (n=263) &   7  &  0.23  &  11.72  &  3.70  &  30.04  &  0.60  &  1.89 \\ 
        2b: View four (s) time slots and select one (n=266)	&   5  &  0.29  &  14.82  &  5.21  &  17.36  &  0.34  &  0.96 \\ 
        2c: View four (r) radius and do not select one (n=87) &   5  &  0.58  &  8.69  &  3.38  &  8.70  &  0.58  &  1.48 \\ 
        2d: View nine (s) seat groups and select one (n=262)	&   10  &  0.63  &  20.85  &  6.26  &  15.84  &  0.48  &  1.60 \\ 
        2e: Click next to options (n=257)	&   1  &  0.69  &  11.67  &  3.56  &  1.44  &  0.09  &  0.28 \\ 
        3a: View six (o) options and select one (n=277)	&   7  &  0.85  &  34.28  &  11.12  &  8.28  &  0.20  &  0.63 \\ 
        3b: Click next to summary	(n=268) &   1  &  0.74  &  11.87  &  3.81  &  1.36  &  0.08  &  0.26 \\ 
        4: View summary and click Submit (n=269)	&   2  &  1.58  &  19.81  &  6.66  &  1.27  &  0.10  &  0.30 \\ 
     \hline
    \end{tabular}
\end{table}

%
%

\section{Discussion and future work}
\label{sec:discussion}

The big \textit{I} notation can be used to estimate interaction complexity of lo-fi and hi-fi concepts, production-level designs, and deployed applications. 
However, this notation does not provide a time estimate. The purpose of this study is to identify the time for an average interaction step per second. To express the interaction steps per time, we introduce the term \textit{interaction speed}.

This study focuses on a user task where users make selections with fixed preferences. For this type of user task, as measured in a movie ticket booking system, the average interaction speed is 1.05 IS/sec, ranging from 0.18 IS/sec on the lower end to 8.15 IS/sec on the upper end (see Table~\ref{tab:times_for_iterations}).

The data shows that the user's interaction speed increases during the session for V1. The mean interaction speed of the training session was 0.53 IS/sec, and then increased from one attempt (1.02 IS/sec) to five attempts (1.44 IS/sec). A possible explanation is that users learned how to use the user interfaces more efficiently with each attempt, thereby increasing their interaction speed.

The mean interaction speed of version 2 (0.66 IS/sec) is significantly lower than the interaction speed of version 1 ($\geq$ 1.05 IS/sec). A possible explanation is that version 2 has more content and options on one user interface than the user interfaces of version 1, which may cause an increased cognitive load \cite{Sweller1994-CognitiveLoadTheory-article}, leading to a reduced interaction speed. We are still surprised by its magnitude, and more research should be performed to understand the exact reasons.

Moving forward, the mean interaction speed of 1.05 IS/sec can be used as an initial hypothesis to translate the interaction complexity, determined with the big \textit{I} notation, into an estimated interaction time. However, more research is needed to determine interaction speeds for other types of user tasks to validate and potentially refine the mean interaction speed.

We can compare the measured interaction speeds with the ones based KLM execution times (see Table~\ref{tab:resulttable4}). The mean KLM interaction speed is approximately 1.34 IS/sec for V1, which is above the interaction speed for V1 (1 attempt) (1.02 IS/sec), and below the interaction speed for V1 (5 attempts) (1.44 IS/sec). Additionally, KLM does not account for learning, one of its known limitations \cite[p. 227]{Olson1990-KLM-HCI-article}. Also, KLM has a known limitation in considering the time for cognitive processes. This becomes evident when comparing the measured interaction speed for V2 (0.66 IS/sec) with the KLM interaction speed for V2 (1.56 IS/sec).

\begin{table}
    \centering
    \label{tab:resulttable4}
    \caption{KLM interaction speeds}
	\begin{tabular}{>{\raggedright\arraybackslash}m{4cm} >{\raggedright\arraybackslash}m{2cm} >{\centering\arraybackslash}m{3cm} >{\centering\arraybackslash}m{2cm}} 
       Version              &         & KLM      & KLM \\
       (Number of attempts) & IS                 & execution time (sec) & IS / sec\\
       \hline
       Version 1 (1)        &   43    & 32.76    & 1.31 \\
       Version 1 (2)        &   75    & 56.20    & 1.33 \\
       Version 1 (3)        &   107   & 79.64    & 1.34 \\
       Version 1 (4)        &   139   & 103.08   & 1.35 \\
       Version 1 (5)        &   171   & 126.52   & 1.35 \\
      \hline
       Version 2 (1)        &   46    & 29.57    & 1.56 \\
      \hline
    \end{tabular}
\end{table}

A limitation of the study is the limited scope of the evaluated user task. The selected user task of purchasing a movie ticket is just one type of user task. If a user task is more exploratory (e.g., find a movie that my friend likes), different interaction speeds can be expected due to prolonged cognitive processes.

Besides the mentioned future work, we also plan to address the limitation of considering cognitive processes. It might be reasonable to identify user task types (decision making, selection, exploration, creation, communication, negotiation, etc.) and measure their interaction speeds. Potentially, a mean time for a cognitive process, or different types of cognitive processes, can be determined. 

%
%
\begin{credits}
\subsubsection{\ackname} This study was funded by the Argentinian National Agency for Scientific and Technical Promotion (ANPCyT), grant number PICT-2019-02485 and by Helmut Degen.

\subsubsection{\discintname}
The authors have no competing interests to declare that are relevant to the content of this article.
\end{credits}

%
%
\bibliographystyle{splncs04}
\bibliography{references}

\end{document}